\documentclass{appolb}
\usepackage{epsfig}

\begin{document}

 \title{%
 BROWNIAN DYNAMICS SIMULATIONS OF SINGLE-FILE MOTION THROUGH NANOCHANNELS
 }

\author{I. D. Kosi\'nska and A. Fuli\'nski
\address{M. Smoluchowski Institute of Physics, Jagiellonian University,
 Reymonta 4, PL-30-059 Krak\'ow, Poland}
}

 \maketitle

 \begin{abstract}

 Algorithm is constructed which models single-file motion of particles
 interacting with each other and with the surroundings. As an example, we
 present the results of Brownian Dynamics simulations of the motion of cations
 moving through a short very narrow channel containing a device called
 ``gate'', which may open and close the channel.

 \end{abstract}

 \section{Introduction}

 Nanochannel transport and physical mechanisms of its regulation are among
 leading open problems in nanoscience. Its importance results from the fact
 that controlled and selective flow of matter through proteins in the cell
 membrane -- achieved by active and passive channels \cite{Hille} -- is one of
 most important biophysical processes in living cells. On the other hand,
 similar functions may be performed by synthetic nanopores which also can
 rectify the ionic currents \cite{{S1},{S2},{fks1}} and pump the ions against
 their concentration gradients \cite{SFpump} and therefore may be used as
 simple models of biological (protein) channels, and, on the other hand, may
 serve as devices for manipulating the transport in the nanoscale. Therefore
 it is important to understand the conditions and properties of material
 transport inside the nanopore.

 The well-known and rather obvious property of the transport of material
 through very narrow pores is that the particles (ions, molecules ...) can
 pass through such channels in the form of single file only \cite{single}.

 In the absence of noise (i. e., in standard Molecular Dynamics simulations)
 time increments $\delta t$ can be made arbitrarily small. This feature makes
 easy (in principle, at least) to keep all particles in prescribed unchanged
 order. In the Brownian Dynamics (BD) the action of random forces may result
 in arbitrarily high velocities and arbitrarily long jumps, time increment
 being irrelevant in this respect. Therefore it is impossible to keep
 particles in still the same ordering by reducing time increments, the more
 that in the presence of noise the time increments cannot be arbitrary
 \cite{{SS},{Moy}}. Some additional procedures are needed.

 We present here the developed by us algorithm which models single-file motion
 of particles interacting with each other and with the surroundings, moving in
 a short very narrow channel containing a device called ``gate'', which may
 open and close the channel. To be specific, we shall discuss in this paper
 the electrostatic and hard-sphere interactions, though the formulas and the
 algorithms themselves can be easily adapted to any (sensible) form of
 interactions.

 \section{The model}

 We use the simplified model which does not take into account the details of
 the channel's structure. Full MD simulations of a K$^+$-channel, including
 its molecular structure, water inside, all ions in the immediate vicinity,
 etc., requires use of total number of atoms in the simulation system above
 $4\times10^4$, and time-steps 0.2 fs \cite{{BR},{SCW}}. Such simulations have
 also some other drawbacks \cite{Levitt}.

 Little is known about the details of the gating mechanism, the more that the
 motions of dangling ends \cite{SFnoise} in synthetic pores are probably quite
 different from the motions of the subunits of proteins constituting the
 biological channels. Therefore, without entering into details of equations of
 motion for the channel's walls, we model the gating process by introducing
 inside the channel the artificial device called ``gate'' which can either
 allow or prevent the flow of particles through the channel.

  The main assumptions are:

 (i) We simulate the motions of the particles inside the {\it simulation zone}
 (SZ) of the lenght $L$, narrow enough to force the particles inside SZ to
 move in the single-file order. Knowledge of the detailed shape (e. g.
 cylinder, cone, hour-glass) is not necessary from this point of view. Regions
 outside are treated as reservoirs for particles both outcoming from and
 ingoing into SZ.

 (ii) We neglect the motions in radial directions, and describe the particles
 as moving along the $z$-axis of the SZ only (quasi-onedimensional motion).
 However, the physical system (electrostatic interactions, etc.) remains
 three-dimensional.

 (iii) The opening and closing of the channel (so-called gating process) is
 modeled by the presence of the charged ``gate'' located inside SZ. The state
 of the gate is determined by its Brownian motion (Wiener process of intensity
 $Q_b$), and by electrostatic interactions with the ions inside SZ and with
 external electric field. The gate opens when the net force exceeds some
 threshold value, and closes otherwise. Minimal approach distance between particle
 and gate is $d_{cg}$.

 (iv) The real channels exhibiting the flicker noise are asymmetric and
 charged. We model these properties by the mentioned above gate, and by
 additional charges located outside SZ.

 (v) Water molecules are not modeled explicitly but are described
 electrostatically by an effective dielectric constant and as the source of
 friction and noise -- as is frequently done \cite{Nadler}.

 No periodic boundary conditions are imposed. Instead,
 in our simulations we assumed (when other rules are satisfied) that

 (i) Particles can leave and enter the simulation zone (SZ) through both
 apertures.

 (ii) Particle leaves the simulation zone (and can be counted to the current
 balance at the given aperture) when its center-of-mass position is smaller
 than the lower threshold, or greater then the higher threshold. In our case
 we accepted as thresholds the particle diameter $d_c$ and SZ length $L$
 minus $d_c$.

 (iii) Single-file assumption implies that when one particle leaves the
 simulation zone, another cannot enter through the same aperture in the same
 time (i. e., during the same time-step).

 (iv) When rule (iii) allows, particle may enter SZ when nearest
 particle is farther that the prescribed smallest distance. In our case the
 smallest distance is $d_c + \epsilon$ ($\epsilon = 0.00001$ nm).

 (v) Particles enter SZ with prescribed finite probabilities $P(0)$ and
 $P(L)$, which may be different for different apertures (i. e. at $x=0$ and
 $x=L$). The probabilities of entrance simulate concentrations outside SZ --
 the lower concentration, the lower probability.

 In our simulations we assumed that (when other rules are satisfied) during
 one time-step only one particle may enter the SZ through a given entrance,
 and, when entering, that it is located at the distance $d_c$ from the
 aperture. This rule can be changed.

The Langevin-type equations of motion for the particles (cations)
moving along the channel reads:
\begin{eqnarray}
 m_i\dot v_i &=& -\gamma_i v_i + R_i(z_i) + F_i(z_i)\,,\nonumber\\
 \dot z_i &=& v_i\,,
 \label{(1)}
\end{eqnarray}
where $v_i$ is the velocity of $i$-th ion, $z_i$ -- the
position, $m_i$ -- the mass, $\gamma_i$ -- the friction
coefficient, $F_i(z_i)$ -- sum of deterministic forces, and
$R_i(z_i)$ -- the random force.

The gate is charged to prescribed value $q_{g}=Z_{g}e$, where
$Z_{g}$ is the valence and can be in two states: open and closed,
respectively. In our simulation important is the absolute value of
the force $F_g$ acting on the gate. We assume $F_g$ to be sum of
deterministic and random forces described below.

 The deterministic force $F_i(z_i)$ experienced by the cations and the gate consist
 of the applied external force (voltage), and the internal Coulomb force
 from other charges.
 The Coulomb interaction between two ions is modified by the
 addition of a short-range repulsive $1/r^{10}$ force, where $r$ is the ion–-ion distance \cite{Moy}.

The random force $R_i$ acting on ions is assumed to be the thermal noise represented
 by the Gaussian white noise. On the other hand the random force experienced by the gate $R_{g}$
 is given by the Wiener process (gate's Brownian motion)
 $R_{g}=\sum_i R_i$.

 In the Brownian Dynamics calculations, $\delta t$ should be of the order of
 $m/\gamma$ \cite{{Kuyucak},{SS}, {FD}}.
Using the Euler scheme

 \begin{equation}
 m\dot v(t) + \gamma v(t) = F(t) \ \to \\
 m\frac{v(t+\delta t) - v(t)} {\delta t} + \gamma v(t) = F(t)
 \label{dis0}
 \end{equation}
would lead to obviously wrong result: $v(t+\delta t) = F(t)$.
Therefore we use the following scheme of discretization:

\begin{eqnarray}
 m\frac{v(t+\delta t) - v(t)} {\delta t} + \frac{\gamma}{2} [v(t+\delta t)
 +  v(t)] &=& F(t)\nonumber\\
\frac{z(t+\delta t)-z(t))}{\delta t} &=& v(t+\delta t).
 \label{dis1}
 \end{eqnarray}
 This computational scheme is similar, though not identical, with that
 described recently in ref. \cite{{SS}, {FD}}.
The ``forward evaluation'' (Eq.3) has stability and
 accuracy implications, and \cite{FD} suggest using it for each extrapolative force calculations.


 %
 \section{Numerical results}

 The length of the simulation zone is $L=10$ nm. This corresponds to
 the real length of biological channels, and -- roughly -- to the length of
 the narrow part of the synthetic channel reported in \cite{SFnoise}

 The net flow of particles through the channel (simulation zone)  was
 calculated either by keeping the balance of particles entering
 and leaving both apertures, or by counting the particles passing the gate in
 both directions. Both procedures lead to the same results.

 Initial values of velocities of particles were drawn from the Maxwell
 distribution with the variance $k_BT/m_c$.
 The results are insensitive on the exact values of temperature
 and mass within rather wide range of temperatures and masses.

A list of the parameters used in the BD simulations is given
 below:
 \medskip

Temperature: $T = 298$ K and $k_BT=4,12\times 10^{-21}$ J,

Mass: $m_c=6.5\times 10^{-26}$ kg, \quad Friction constant:
$\gamma_c = 2.08\times 10^{-12}$ kg/s,

Dielectric constant: $\epsilon_{w} = 81$,\quad Voltage:
$U=1.77\times 10^{-2}$ V

Ion diameter: $d_c = 0.266\times 10^{-9}$ m,\quad Ion-gate
min.distance: $d_{cg} = 2.5 d_c$,

Valences: $Z_{c}=+1$,\quad $Z_{g}=-50$,

Intensity of short-range force: $F^0_{SR}= 444 \times 10^{-9}$ N,

Intensity of noise: $Q_{c}=0.47\times 10^{-9}$ N,\quad$Q_{g}=0.01
Q_i$.

 Intensity of the gate's noise $Q_g$ is different from cations' one $Q_c$
 (and is taken as a free parameter) due to the difference of masses, and also
 due to a kind of ``stiffness'' of (or hindrances in) the motions of channel's
 walls constituents.
\begin{figure}[ht]
    \begin{center}
      \includegraphics[width=6cm, height=5cm]{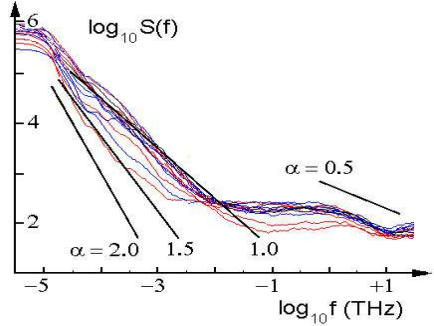}
       \caption{\small{Power spectra $S(f)$ of the stochastic series of
subsequent values of the net number of cations $m_n$ leaving the
simulation zone. Red: $S(f)$ for 7 different realizations of the
intrinsic noises, the same values of all parameters in every
series: $\epsilon = 81$, $m_c = 6.5 \times 10^{-26}$ kg, $U=
17.7*10^{-3}$ V, $\delta t = 31\times 10^{-15}$s, $Q_g = 0.01
Q_c$, $F^0_{SR} = 444 \times 10^{-9}$ N, gate thresholds =
$\pm1100 \times 10^{-12}$ N. Blue: $S(f)$ with the same
realizations of the intrinsic noises, for 7 different values of
all parameters in every series. In every series only one parameter
is changed: $\epsilon = 0.93 \epsilon^0$, $m = 0.77 m^0$, $U =
1.33 U^0$, $\delta t = 0.8 \delta t^0$, $Q_g = 0.75 Q_g^0$,
$F_{SR} = 0.6 F^0_{SR}$, where $p^0$ denotes the value of the
given parameter from the panel A.}}
    \end{center}
    \label{fig:1}
\end{figure}
 In all simulations first $10^6$ steps were rejected. The power spectrum was
 calculated from runs of length $10^7\delta t$. The power spectrum of the series
$\{m_1, \dots m_N \}$ is
\begin{equation}
S(f) = \frac{1}{N} \bigg\vert \sum_{n=1}^N m_n e^{-2\pi ifn} \bigg\vert^2\,,
\label{spectrum}
\end{equation}
where $m_n$ denotes either the net number of particles leaving SZ
during the $n$-th step (then $m_n$ can be either positive, zero,
or negative), the number of particles inside SZ at the end of the
$n$-th step ($m_n = N_p \ge 0$), or the state of the gate during
the $n$-th step (then $m_n =\{0,1\}$). All these power spectra are
dimensionless.
\begin{figure}[ht]
    \begin{center}
      \includegraphics[width=6cm, height = 5cm]{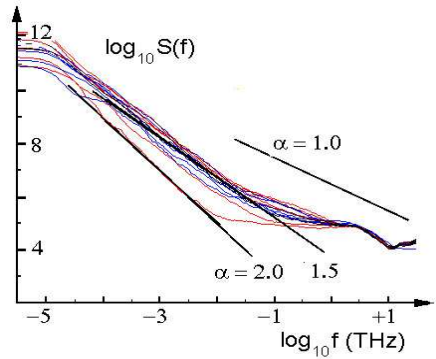}
       \caption{\small{Power spectra $S(f)$ of the stochastic series of subsequent values of the
number of cations $N_p$ inside the simulation zone. Notation and
values of parameters the same as in Fig. 1.}}
    \end{center}
    \label{fig:2}
\end{figure}
 There are data that suggest that inside very narrow pores the physical
 properties of aqueous solutions, such as dielectric constant, density,
 diffusion coefficient, viscosity, solvatation of ions (i. e., their effective
 diameters), etc. may differ from their bulk values \cite{data}. Therefore we
 checked how the changes of such parameters influence our model. We found that
 the quantitative changes of calculated values of net currents and of
 frequency spectra resulting from reasonable variations of these parameters
 are within the limits of quantitative differences resulting from different
 realizations of the noise. The results are shown in Figs.1--3.
 These observations suggest robustness of the model.
\begin{figure}[ht]
    \begin{center}
      \includegraphics[width=6cm, height=5cm]{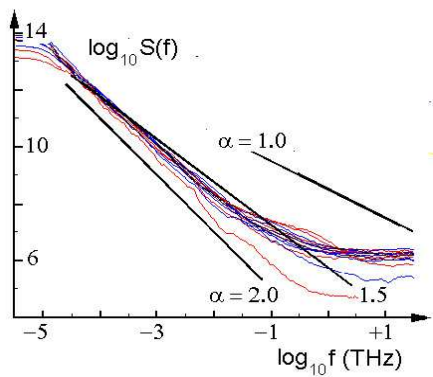}
       \caption{\small{Power spectra $S(f)$ of the stochastic series of
subsequent values of the state of the gate. Notation and values of
parameters the same as in Fig. 1.}}
    \end{center}
    \label{fig:3}
\end{figure}
 On the other hand, the model is sensitive with respect to the changes of
 relative strength of random and deterministic forces -- decrease of the
 dielectric constant with noise unchanged, or increase of noise with
 electrostatic forces unchanged changed significantly the results. E. g.,
 either too strong gate noise or too strong electrostatic force (i. e., low
 dielectric constant) dampen the flicker noise.

 When the single-file limitations are removed, all the power spectra shown in
 Figs. 1-3 become $S(f) \sim f^{-1.5}$, i .e, the corresponding processes
 behave like the Wiener process.

 %
%
%
%

%
 \section{Appendix}

 Here we present the codes for the single-file motion. The codes for entrances
 and exits of particles, for the number of particles located to the left of
 the gate, as well as the codes for the determination of the state of the gate
 (open or closed), and for the equations of motion are standard and will not
 be reproduced here.

 Single-file procedures are based on the fact that the given particle (cation)
 $i$ cannot move farther that its neighbours $i-1$ and $i+1$, which in turn
 are limited by their neighbours, $i$ and $i-2$ or $i+2$, etc. Therefore their
 positions need to be recalculated. In the simplest version, it is assumed
 that particles meet at the middle of their former positions. In the better
 versions such a pair of particles meets at the position calculated from their
 former positions and from their new velocities. On the other hand, the
 particles retain their velocities until a given pair meets, then they collide
 and -- in the simplest version -- exchange their velocities (behave as hard
 spheres). Again, it is possible to refine this simplest procedure. Because
 the results of the above-described procedure depend on whether the
 recalculations are done``up'' or ``down'', i.e., from particle number 1 to
 $N$, or from $N$ to 1, both reorderings are realized independently, their
 results are averaged, and the whole scheme is iterated until
 self-consistency is attained.

 Before using the SFM-codes below, one needs to supply the values of entries
 of three main arrays: ZK[Nkmax], VK[Nkmax], in which the positions and
 velocities of particles inside the simulation zone are stored, ZK0[Nkmax]
in which former positions are remembered, and two
 auxiliary ones: ZKG[Nkmax] and ZKH[Nkmax] for storing intermediate data.
 It is needless to say that these arrays should be declared as external
 variables.

 Nkmax denotes here the maximal, Nk (in the codes) -- the actual number of
 particles inside the simulation zone.

 We present here separate single-file codes for open and for closed channel.
 Before calling the  S-F code for a closed channel, one needs to
 calculate the number of particles located to the left of the gate, denoted
 in the codes as Nkgl.

 \bigskip

 {\tt

\qquad \qquad  //  single-file ordering:  open channel

nrep = 0; repeat = 1;

while(repeat == 1 \&\& Nk > 1)

\{

\quad repeat = 0; nrep++;

\quad orderlow(1,Nk,dcc);                   // pairs (1,2),...(Nk-1,Nk)

\quad orderup(0,Nk-1,Nk,dcc);             // pairs (Nk,Nk-1),...(2,1)

\quad ave(1,Nk);

\quad subst(Nk);

\}

\medskip
\qquad \qquad   // single-file ordering:  closed channel

while(repeat == 1)

\{

\quad repeat = 0; nrep++;

\qquad \qquad // ordering to the right of the gate:

\quad if(Nkgl < Nk)

\quad \{

\quad \quad ic = ordergr(bp,Nkgl,Nkgl+1,Nk,dcc); \qquad   // at the gate

\quad \quad if(ic < Nk)   \qquad    // remaining particles

\quad \quad \{

\quad \quad \quad orderlow(ic,Nk,dcc); \qquad   // pairs ic,ic+1),...(Nk-1,Nk)

\quad \quad \quad orderup(0,Nk-ic,Nk,dcc); \qquad //pairs Nk,Nk-1),...(ic+1,ic)

\quad \quad \quad ave(ic,Nk);

\quad \quad \}

\qquad \qquad // ordering to the left of the gate:

\quad \}

\quad ic = ordergl(bl,Nkgl,0,Nkgl,dcc);  \qquad // at the gate

\quad if(ic > 1) \qquad      // remaining particles

\quad \{

\quad \quad orderlow(1,ic,dcc);\qquad   // pairs (1,2),...(ic-1,ic)

\quad \quad orderup(0,ic-1,ic,dcc);   \qquad // pairs  (ic-1,ic),...,(2,1)

\quad \quad ave(1,ic);

\quad \}

\quad subst(Nk);

\}

\quad

void orderlow(int m, int N, double d)

\{

\quad int i;

\quad for(i=m;i<N;i++)

\quad \{

\quad \quad if(ZK[i] > ZK[i+1] - d)

\quad \quad \{

\quad \quad \quad ZKH[i] = 0.5*(ZK0[i] + ZK0[i+1] - d);

\quad \quad \quad ZKH[i+1] = ZKH[i] + d + 0.00001;

\quad \quad \quad VKH[i] = VK[i+1]; VKH[i+1] = VK[i];

\quad \quad \quad repeat = 1;

\quad \quad \}

\quad \}

\quad return;

\}

\quad

void orderup(int m, int N, int M, double d)

\{

\quad int i, j;

\quad for(j=m;j<N;j++)

\quad \{ i = M - j;

\quad \quad if(ZK[i] < ZK[i-1] + d)

\quad \quad \{

\quad \quad \quad ZKG[i] = 0.5*(ZK0[i] + ZK0[i-1] + d);

\quad \quad \quad ZKG[i-1] = ZKG[i] - d - 0.00001;

\quad \quad \quad VKG[i] = VK[i-1]; VKG[i-1] = VK[i];

\quad \quad \quad repeat = 1;

\quad \quad \}

\quad \}

return ;

\}

\quad

int ordergl(double b, int Nkgl, int m, int N, double d)

\{

\quad int i, j, ii;

\quad double a, c;

\quad bc = b; ii = Nkgl;                   // bc = gate + dcd

\quad for(j=m;j<N;j++)

\quad \{

\quad \quad i = Nkgl - j;

\quad \quad if(ZK[i] > bc)

\quad \quad \{

\quad \quad \quad ZK[i] = bc; ZKG[i] = bc; ZKH[i] = bc;

\quad \quad \quad bc -= d; ii = i;

\quad \quad \quad if(i == Nkgl) VK[i] = -VK[i]; else

\quad \quad \quad \{

\quad \quad \quad \quad a = VK[i]; c = VK[i+1];

\quad \quad \quad \quad if(fabs(a) > fabs(c))

\quad \quad \quad \quad \{

\quad \quad \quad \quad \quad VK[i] = a + c; VK[i+1] = 0;

\quad \quad \quad \quad \}

\quad \quad \quad \quad else

\quad \quad \quad \quad \{

\quad \quad \quad \quad \quad VK[i+1] = a + c; VK[i] = 0;

\quad \quad \quad \quad \}

\quad \quad \quad \}

\quad \quad \quad repeat = 1;

\quad \quad \}

\quad \quad else break;

\quad \}

\quad return ii;

\}

\quad

int ordergr(double b, int Nkgl, int m, int N, double d)

\{

\quad int i, ii;

\quad double a, c;

\quad bc = b; ii = Nkgl + 1;

\quad for(i=m;i<=N;i++)

\quad \{

\quad \quad if(ZK[i] < bc)

\quad \quad \{

\quad \quad \quad ZK[i] = bc; ZKG[i] = bc; ZKH[i] = bc;

\quad \quad \quad bc += d; ii = i;

\quad \quad \quad if(i == Nkgl+1) VK[i] = -VK[i]; else

\quad \quad \quad \{

\quad \quad \quad \quad a = VK[i]; c = VK[i-1];

\quad \quad \quad \quad if(fabs(a) > fabs(c))

\quad \quad \quad \quad \{

\quad \quad \quad \quad \quad VK[i] = a + c; VK[i-1] = 0;

\quad \quad \quad \quad \}

\quad \quad \quad \quad else

\quad \quad \quad \quad \{

\quad \quad \quad \quad \quad VK[i-1] = a + c; VK[i] = 0;

\quad \quad \quad \quad \}

\quad \quad \quad \}

\quad \quad \}

\quad \quad else break;

\quad \} // i

\quad return ii;

\}

\quad

void ave(int m, int N)

\{

\quad int i;

\quad for(i=m;i<=N;i++)

\quad \{

\quad \quad ZK[i] = (ZKG[i] + ZKH[i])/2.0;

\quad \quad VK[i] = (VKG[i] + VKH[i])/2.0;

\quad \}

\quad return;

\}

\quad

void subst(int N)

\{

\quad int i;

\quad for(i=1;i<=N;i++)

\quad \{

\quad \quad ZKG[i] = ZK[i]; ZKH[i] = ZK[i];

\quad \quad VKG[i] = VK[i]; VKH[i] = VK[i];

\quad \}

\quad return;

\}

}

\bigskip

 For simplicity, we give here, in the functions {\tt orderlow} and {\tt
 orderup} (lines {\tt ZKH[i] = 0.5(ZK0[i] + ZK0[i+1] - d), ZKG[i] = 0.5(ZK0[i]
 + ZK0[i-1] + d)}) the simplest form of the recalculations of the correct
 positions of pairs of particles as contact positions of the pair in the
 middle of their former positions. These positions can be determined with
 better accuracy by taking into account particles' velocities, and their
 equations of motion as well. The appropriate codes are obvious and are not
 reproduced here.

These codes can be written in a more compact way. The form presented here is
 -- in our opinion -- better legible and more self-explanatory.

 \end{document}